\documentclass[prd,aps,superscriptaddress,showpacs,amssymb,
amsmath,twocolumn]{revtex4}
\usepackage{graphicx}

\newcommand{\edinburgh}{School of Physics, University of
Edinburgh, King's Buildings, Edinburgh EH9 3JZ, U.K.}

\newcommand{\papertype}{Brief Report}

\newcommand{\figwidth}{3.3in}

\newcommand{\naive}{na\"{\i}ve}
\newcommand{\fattad}{\textsc{fat7tad}}
\newcommand{\asqtad}{\textsc{asqtad}}
\newcommand{\npoly}{\ensuremath{n_{\text{poly}}}}
\newcommand{\rloc}{\ensuremath{r_{\text{loc}}}}
\newcommand{\eff}{{\text{eff}}}
\newcommand{\nth}{\ensuremath{n^{\text{th}}}}

\def\slashchar#1{\setbox0=\hbox{$#1$}           
   \dimen0=\wd0                                 
   \setbox1=\hbox{/} \dimen1=\wd1               
   \ifdim\dimen0>\dimen1                        
      \rlap{\hbox to \dimen0{\hfil/\hfil}}      
      #1                                        
   \else                                        
      \rlap{\hbox to \dimen1{\hfil$#1$\hfil}}   
      /                                         
   \fi}                                         %

\begin{document}

\title{The locality of the square-root method for improved staggered quarks}

\author{A. \surname{Hart}}
\author{E. \surname{M\"{u}ller}}
\affiliation{\edinburgh}

\begin{abstract}
  We study the effects of improvement on the locality of
  square-rooted staggered Dirac operators in lattice QCD simulations.
  We find the localisation lengths of the improved operators (\fattad\ 
  and \asqtad) to be very similar to that of the one-link operator
  studied by Bunk et al., being at least the Compton wavelength of the
  lightest particle in the theory, even in the continuum limit. We
  conclude that improvement has no effect.  We discuss the
  implications of this result for the locality of the \nth-rooted
  fermion determinant used to reduce the number of sea quark flavours,
  and for possible staggered valence quark formulations.
\end{abstract}

\preprint{Edinburgh 2004-10}


\maketitle

\section{Introduction}
\label{sec_introduction}

One of the biggest headaches in lattice QCD calculations is the
inclusion of quarks in a realistic and readily simulated manner.
Staggered fermions have the advantage over Wilson--like quarks that
their residual chiral symmetry protects the light quark mass from
additive renormalisation. Staggered fermions are also much cheaper to
simulate than those from the overlap formulation. The disadvantage is
that the formalism naturally yields $N_f = 4$ degenerate fermion
species (known as ``tastes''). In the continuum limit these tastes
decouple, but at finite lattice spacing, $a$, there are
$\mathcal{O}(a^2)$ taste-changing interactions. Improvement of the
action can systematically reduce these interactions, leading to
smaller splittings in hadron mass multiplets
\cite{Follana:2003},
a more physical Dirac spectrum
\cite{Follana:2004sz,Durr:2004as}
and the correct chiral suppression of the topological susceptibility
\cite{Hasenfratz:2001wd,Bernard:2003gq}.

True QCD, however, requires a light ``$2+1$''-flavour sea quark mass
degeneracy.  The usual way of tackling this is to replace the fermion
determinant in the partition function by its \nth-root, with $n$
either~2 or~4, to give reduced-taste staggered fermions
\cite{Fucito:1981fh,Marinari:1981qf}.
This approach is inspired by the continuum factorisation into
independent tastes. It is not certain, however, whether this process
yields a theory that is sufficiently local that we are confident of
the universality in the continuum limit, and will arrive at the
correct theory, QCD, as $a \to 0$.

To answer this we seek to prove at least the existence of a local
operator whose determinant matches that of the reduced-taste sea
quarks. Explicitly formulating this operator would also provide a
consistent description of the valence quarks --- currently lacking ---
and allow an unambiguous perturbative analysis of the theory.
Attention so far has focused on the simplest candidate, the
\nth-rooted Dirac operator, as $( \det \slashchar{D} )^{1/n} = \det
(\slashchar{D}^{1/n})$.

The square-root of the simplest, one-link staggered operator was
studied in
\cite{Bunk:2004br}.
Free fermions were shown to be non-local both at finite $a$ and in the
continuum limit, with a localisation length of order the inverse of
the quark mass, $m$. The limits of the eigenvalue spectrum placed a
similar upper bound on the interacting theory. Numerical measurements
showed the operator had a continuum limit localisation length at least
as large as the correlation length of the lightest confined state, the
Goldstone pion.

In this \papertype\ we investigate to what extent the non-locality (or
otherwise) is due to the taste-changing interactions. Using a similar
methodology to
\cite{Bunk:2004br},
we compare the locality of the \fattad\ and \asqtad\ improved
staggered operators at various lattice spacings. We show the size of
the taste-changing interactions has almost no effect, and all the
square-rooted staggered operators are non-local in the continuum limit
on a scale at least that of the largest Compton wavelength. We discuss
the implications of this for possible reduced-taste formulations for
valence and sea quarks in the context of the expected properties of
the infrared Dirac spectrum.

\section{Method}
\label{sec_meas}

An exponential locality length, \rloc, can be defined as follows
\cite{Bunk:2004br}.
Given a candidate Dirac operator $\slashchar{D}(x,y)$, we apply it to
a point source $\xi(y) = \delta_{yz}$ to define a wavefunction
$\psi(x) = \sum_y \slashchar{D}(x,y) \xi(y)$. The operator is
exponentially local if the locality function $f(r)$ is bounded by
\begin{eqnarray}
f(r) \equiv  \max_{||x-z|| = r} \left\{ \left| \psi(x) \right| \right\} 
\leqslant \exp \left( - \frac{r}{\rloc} \right) \; .
\label{eqn_def_locality}
\end{eqnarray}
We require $\rloc \to 0$ in the continuum limit to guarantee
universality, as has been seen for overlap fermions
\cite{Hernandez:1998et}.

We measure Eqn.~(\ref{eqn_def_locality}) for the one-link, \fattad\ 
and \asqtad\ staggered quark formulations
\cite{Orginos:1998ue,Orginos:1999cr}.
The lattices studied are described in Table~\ref{tab_params}. We use
periodic boundary conditions and study three lattice spacings,
choosing the quark mass such that the pion (lightest singlet
pseudoscalar meson) mass is fixed in physical units. For compatibility
with
\cite{Bunk:2004br},
we choose $m$ such that $r_0 m_\pi \simeq 1.3$ (or $520~\text{MeV}$)
\cite{Gupta:1991mr,Kim:1995tg,Kim:1999ur}.
The lattice size is fixed around $1.5~\text{fm}$, or four Compton
wavelengths of the pion.

We use the same $m$ for calculations with the improved fermion
formulations. Whilst the different renormalisations will change the
value of the pion mass, Ref.
\cite{Orginos:1998ue}
shows this effect is only slight, and does not affect the conclusions
of this study.

If $M$ is the staggered Dirac operator, then $M^\dagger M$ defined on
the even sites is an Hermitian, positive-definite matrix representing
$N_f = 4$ tastes
\cite{Polonyi:1984zt,Martin:1985yn}.
We study the locality of $\slashchar{D} = \sqrt{M^\dagger M}$ using
Eqn.~(\ref{eqn_def_locality}). In this, $| \cdot |$ is the SU(3)
colour norm, and $|| \cdot ||$ the periodic $L_1$ (``taxi driver'')
norm
\cite{Bunk:2004br}.

The square-root function is approximated using a Tchebyshev polynomial
\cite{numrec}
of order $\npoly=500$.  The analytic upper bound on the truncation
error is $\mathcal{O}(10^{-2})$
\cite{Kennedy:2004tj}.
Before discussing the results, we
must first be sure that the truncation errors in the polynomial do not
affect our estimates of the locality function.

\begin{table}[t]

\caption{
\label{tab_params} The Wilson gauge action ensembles studied. 
The $\beta=6.0$ configurations are from
\cite{Kilcup:1997hp}. 
Tadpole improvement factors $u_0$ come from the mean plaquette
\cite{Lucini:2001ej}. We set the scale using $r_0$ \cite{Necco:2001xg}.}

\begin{ruledtabular}
\begin{tabular}{ccccccc}
  $\beta$ & $L^3T$ & $N_{\mathrm{conf}}$ & $r_0$ & 
  $aL/\text{fm}$ & $u_0$ & $m$ \\
\hline
  5.8 & $12^4$          & 107 & 3.673~(5) & 1.63 & 0.8680 & 0.017 \\
  6.0 & $16^3 32$       & 221 & 5.371~(15) & 1.49 & 0.8778 & 0.010 \\
  6.2 & $24^4$          & 94 & 7.380~(26) & 1.63 & 0.8851 & 0.007\\
\end{tabular}
\end{ruledtabular}
\end{table}
\begin{figure}[b]
\includegraphics[width=\figwidth,clip]{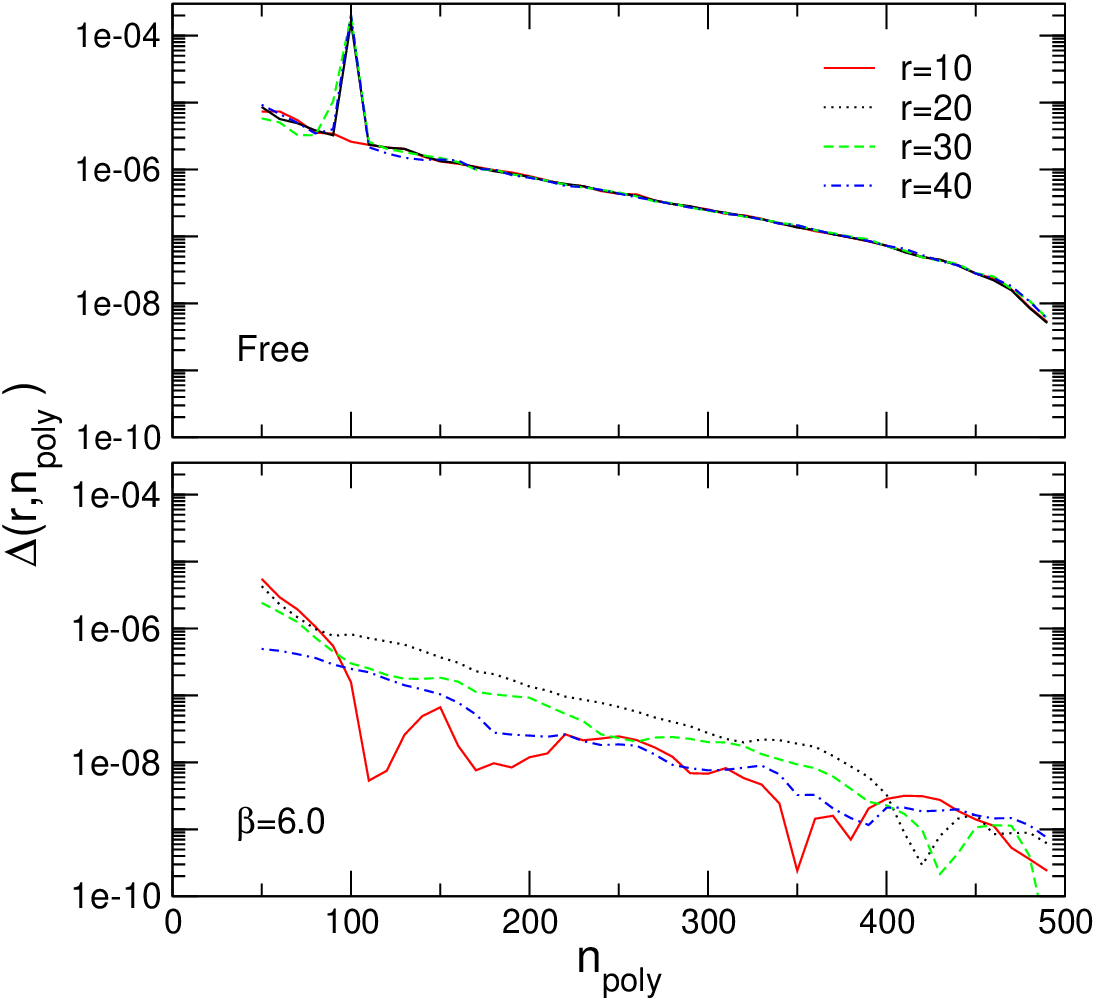}
\caption{\label{fig_accuracy} Relative accuracy of the localisation function 
  using the polynomial approximation to the square root of the
  one-link staggered action. Measurements are on $16^3 32$ for the
  free theory and $\beta=6.0$.  }
\end{figure}

Measuring $f(r)$ for various $\npoly \leqslant 500$,
the error should decrease exponentially, and so
\begin{eqnarray}
\Delta(r,\npoly) & \equiv & 
\left| f(r,\npoly) \right. - \left. f(r,500) \right| 
\nonumber \\
& = & A e^{-B \npoly } \left(
1 - e^{-B \left( 500 - \npoly \right) } \right)
\nonumber \\
& \simeq & A e^{-B \npoly } ~~ \text{for } 0 < \npoly \ll 500 
\end{eqnarray}
and $A(r),B(r) > 0$. We plot $\Delta$ at various fixed $r$ in
Fig.~\ref{fig_accuracy}, for free fermions and for a representative
gauge background at $\beta = 6.0$. Extrapolating the linear region to
$\npoly = 500$, we estimate the truncation error to be
$\mathcal{O}(10^{-8})$. This is an order of magnitude smaller than the
minimum of $f(r)$. As improved fermions should lie somewhere between
the free and the one-link interacting case, we use $\npoly = 500$ in
all our analysis.

\begin{figure}[t]
\includegraphics[width=\figwidth,clip]{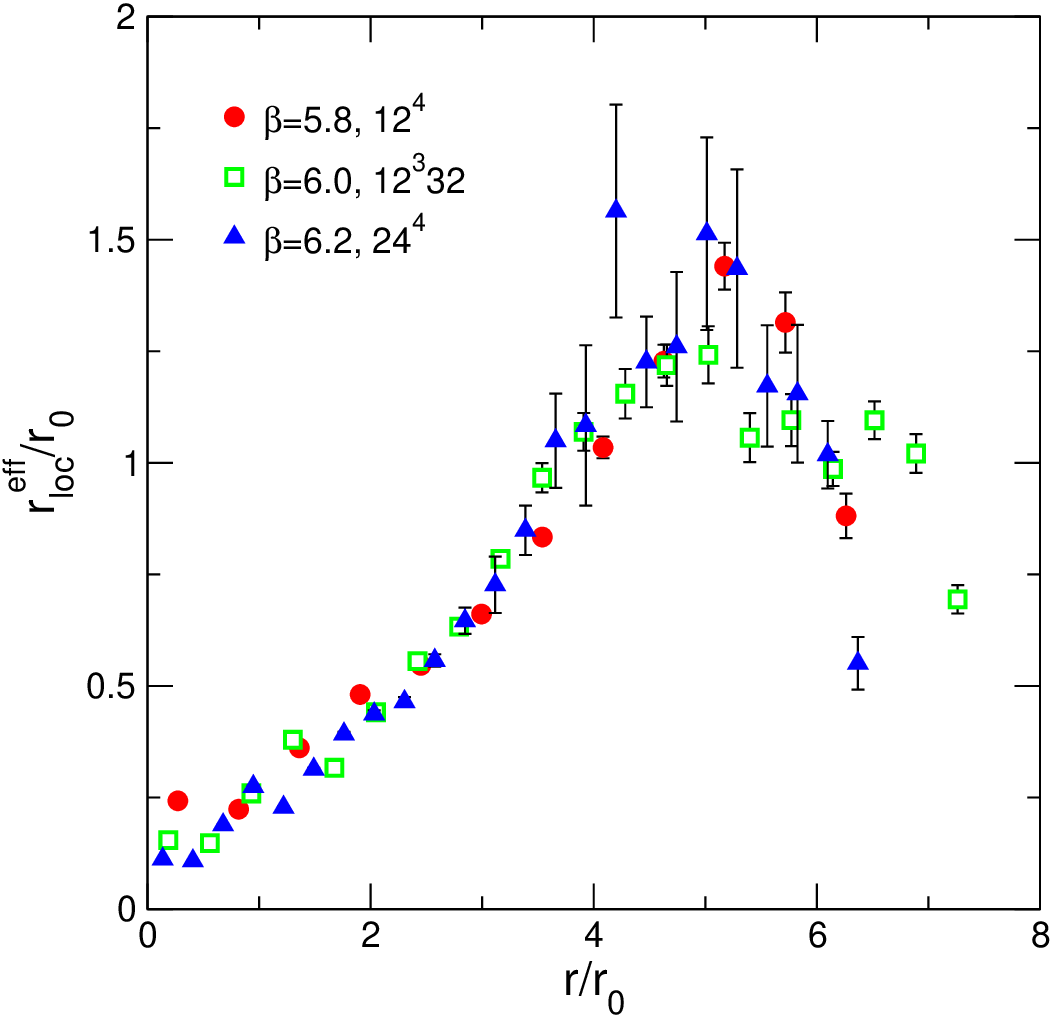}
\caption{\label{fig_rloc_scaling} Scaling of the effective localisation
  length for reduced-flavour $N_f = 2$ \asqtad\ fermions.}
\end{figure}

We may ask why the actual error is so much smaller than the
theoretical bound. A heuristic argument is that the errors on the
Tchebyshev approximation are heavily concentrated at the lower end of
the spectrum. The eigenvalues of the Dirac operator are, by contrast,
relatively sparse here. Assuming the localisation function is not
completely dominated by the infrared modes, it is reasonable that
truncation errors towards the middle of the operator spectrum (which
are many of orders of magnitude smaller) are more representative of
the error on $f(r)$.

\section{Results}

Using the locality function, we define an effective localisation
length $\rloc^\eff(r)$ at distance $r$ by
\begin{equation}
  \rloc^\eff
  \left( \frac{r_i+r_{i+1}}{2} \right) = 
\frac{(r_{i+1}-r_i)}{\log \left( f(r_i)/f(r_{i+1})\right)} 
\label{r_loc_formula}
\end{equation}
for each pair of subsequent distances $r_i$.

We plot this function for the square-rooted \asqtad\ operator in
Fig.~\ref{fig_rloc_scaling}, scaling everything in units of $r_0$.  At
small distances $f(r)$ is almost a power law (with exponent somewhere
between $-3$ and $-4$). Comparing our results with larger volumes $aL
\simeq 2.1~\text{fm}$, only results for $r/r_0 \lesssim 3$ are free of
finite volume effects.  In this range there is no clear evidence for a
crossover to exponential locality, so localisation length estimates
must be treated as lower bounds only.  Nonetheless, these bounds do
not scale to zero with the lattice spacing, but rather remain
constant. It therefore seems highly likely that the actual
localisation length also remains finite in physical units in the
continuum limit.

\begin{figure}[t]
\includegraphics[width=\figwidth,clip]{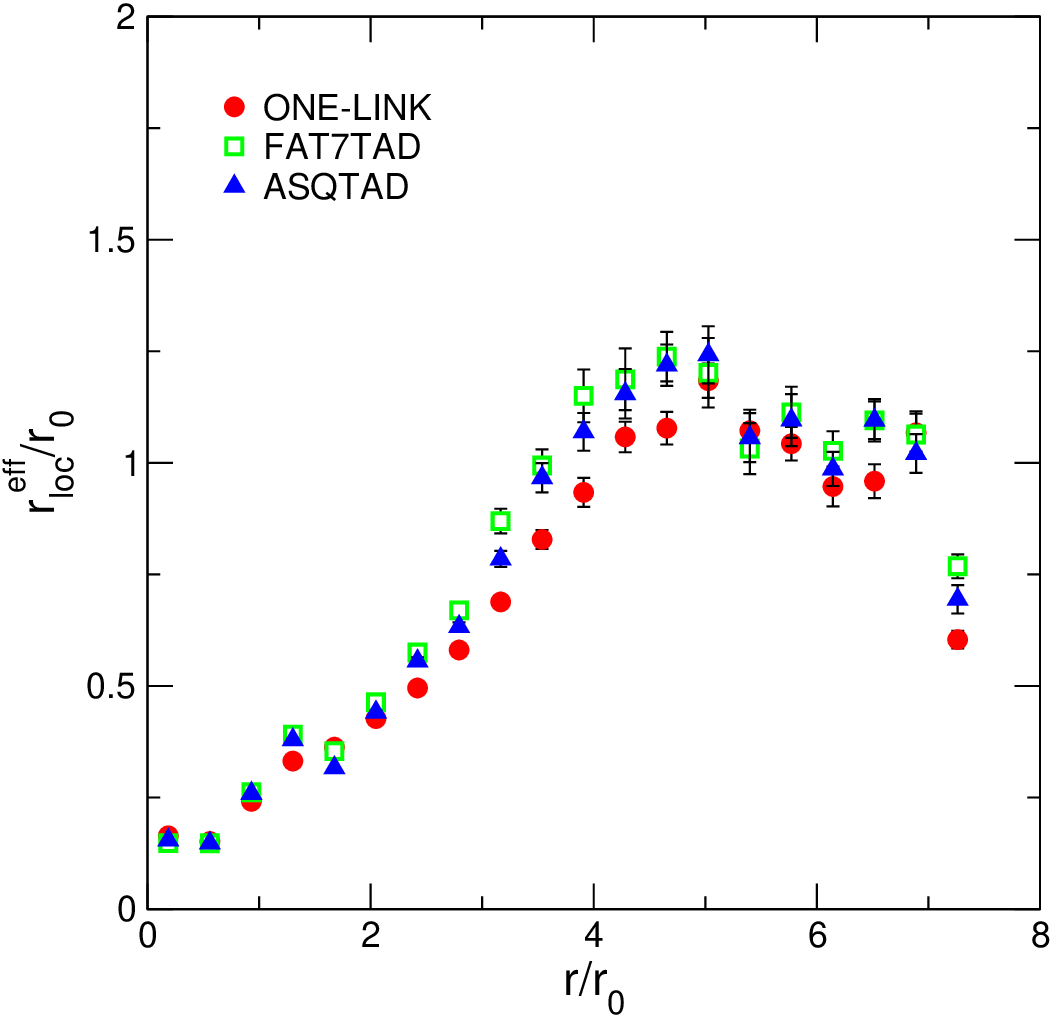}
\caption{\label{fig_rloc_compare} A comparison of the effective 
  locality length for various reduced-flavour $N_f=2$ staggered
  operators at $\beta=6.0$.}
\end{figure}

We conclude, then, that the square-rooted \asqtad\ Dirac operator is
severely non-local in the continuum limit. In
Fig.~\ref{fig_rloc_compare} we compare at fixed lattice spacing the
three operators. The effective locality lengths are extremely similar.
This was also the case at the other lattice spacings. It would
therefore appear that the improvement programme for staggered quarks
has no effect on the locality of the square-rooted Dirac operator.

\section{Discussion}

Understanding how to consistently reduce the number of staggered
tastes for the sea and valence quarks is vital to realistic QCD
simulations. A proposed solution for the sea quarks is to take the
\nth-root of the fermion determinant. The locality (and legitimacy) of
the reduced-taste theory is not clear, but may be answered by
attempting to find a local reduced-taste Dirac operator with the same
determinant. Using this same operator to describe the valence quarks
(rather than the $N_f = 4$ staggered operator used at present) would
also avoid the potential mode mismatches seen for other mixed
valence/sea fermion combinations
\cite{Durr:2003xs}.

We have examined a candidate two-flavour Dirac operator, inspired by
the \nth-root trick used to reduce the number of sea quark tastes. We
have seen that the locality of this square-rooted operator is as bad
for improved staggered formulations (\fattad\ and \asqtad) as it was
for the one-link case
\cite{Bunk:2004br}:
the locality length for $5.8 \leqslant \beta \leqslant 6.2$ shows no
signs that \rloc\ vanishes in the continuum limit. By contrast, \rloc\ 
appears to be at least the Compton wavelength of the lightest
particle.

In many areas improved staggered quarks show a definite improvement
over the one-link case
\cite{Hasenfratz:2001wd,Bernard:2003gq,Follana:2003,Follana:2004sz}.
That there is almost no commensurate change in the locality suggests
that the \nth-rooted operator is not a useful way of formulating the
valence quarks. It is therefore unlikely to be a good choice for
studying the locality of the determinant; a better choice would be an
operator with more physical properties.

Given a sufficiently smooth gauge background of topological charge
$Q$, a lattice Dirac operator should have an infrared spectrum
characteristic of the number of tastes/flavours it represents.  The
Atiyah-Singer Index Theorem predicts that there should be (at least)
$N_f \; |Q|$ near-zero modes
\cite{Smit:1987fn,Follana:2004sz,Durr:2004as}.
The remaining low-lying modes should lie in near-degenerate
$N_f$-plets, with any splitting due to taste-changing interactions
\cite{Follana:2004sz,Durr:2004as}.
The multiplets should follow a universal distribution given, for
instance, by random matrix theory
\cite{Shuryak:1993pi,Follana:2004sz}.
Taking the \nth-root will reduce neither the number of zero modes nor
the multiplicity of other eigenvalues by factors of $n$. In addition,
eigenvalue ratios $\langle \lambda_s^{1/n} \rangle / \langle
\lambda_t^{1/n} \rangle$ can no longer agree with the universal
distribution.

By analogy with the reduction of 16 \naive\ fermion species to 4
ultra-local staggered tastes, it seems likely that a physical
reduced-taste staggered Dirac operator will, if it exists, be obtained
by projection rather than rooting. Unlike \naive\ to staggered,
however, taste-breaking interactions prevent this being exact at
finite lattice spacing, but improved staggered spectra already show
much closer taste symmetry
\cite{Follana:2004sz,Durr:2004as}.
It is not clear whether such a reduced-taste operator would be local
or even renormalisable. It is, however, important to study this to
understand the systematic uncertainty of using the \nth-root method in
precision lattice QCD simulations.

\begin{acknowledgments}
  
  We are grateful to K. Jansen and A.D. Kennedy for useful
  discussions, as well as A.D.K. for a critical reading of the
  manuscript. A.H. thanks the U.K. Royal Society, and E.M. the DAAD
  German Academic Exchange Service for financial support.

\end{acknowledgments}

\bibliographystyle{h-physrev4}
\bibliography{locality_refs}

\end{document}